\documentclass[showpacs,prb,twocolumn,floats,superscriptaddress]{revtex4}

\usepackage{graphicx}
\usepackage{amsmath}
\usepackage{amssymb}
\usepackage[colorlinks=true,citecolor=blue,linkcolor=blue]{hyperref}
\usepackage{amsfonts}
\usepackage[latin9]{inputenc}
\usepackage{color}
\usepackage{textcomp}

\renewcommand{\vec}[1]{\ensuremath{\boldsymbol{#1}}}

\begin{document}

\title{Substrate effects on transport properties of a biased AA-stacked bilayer graphene}
\date{\today }
\author{Hasan M. Abdullah}
\email{alshehab211@gmail.com}
\affiliation{Department of Physics, King Fahd University of Petroleum and Minerals, 31261 Dhahran, Saudi Arabia}
\affiliation{Saudi Center for Theoretical Physics, P.O. Box 32741, Jeddah 21438, Saudi Arabia}

\author{H. Bahlouli}
\affiliation{Department of Physics, King Fahd University of Petroleum and Minerals, 31261 Dhahran, Saudi Arabia}
\affiliation{Saudi Center for Theoretical Physics, P.O. Box 32741, Jeddah 21438, Saudi Arabia}

\begin{abstract}

  The important experimental advances in graphene fabrication and its peculiar transport properties  motivated researchers to utilize graphene as a potential basis for the next generation of fast and smart electronic devices. In this article, we investigate the influence of a potential substrate on the transport properties of a biased AA-stacked $n$-$p$-$n$ bilayer graphene junction (AA-BLG). Using the Dirac Hamiltonian with the transfer matrix approach we obtain the transmission probabilities and thus the respective conductance. In the presence of the  induced mass-term   the energy spectrum and  the intra-cone transmission drastically change while   the  inter-cone transmission remains zero. On the other hand, the bias   slightly alters the energy spectrum but it significantly affects the transport properties due to its ability to \textit{switch on} the inter-cone transmission. In addition, we find that Klein tunnelling is attenuated in the presence of the induced mass-term which can improve the carriers confinement in such configurations.
Our findings provide possible experimental measurements to determine the interlayer coupling and the induced mass terms in graphene bilayer based on  conductance and band structure measurements.\end{abstract}

\maketitle

\section{Introduction}
Since its experimental realization in 2004\cite{Geim_2007}, graphene and its multilayer systems have triggered an avalanche of interest due to its remarkable electronic, optical and  mechanical properties.   Such extraordinary properties make graphene  a promising material for nanoscale device applications in the future such as its potential use  in sensors, detectors, electronics, ...etc\cite{Castro_Neto_2009, Berdiyorov_2016,Tit_2017}. . Bilayer graphene exists with two different types of stacking, namely, AB-(Bernal) and AA-stackings (AB-BLG and AA-BLG).  Due to the stability of AB-BLG, it has been subjected to  considerable  theoretical and experimental investigations\cite{Ohta_2006,Goerbig_2011,Abdullah_2016}.  Albeit the old belief in the instability of AA-BLG, recent stable samples were successfully realized.\cite{Lee_2008,Borysiuk_2011,de_Andres_2008,Liu_2009}.
Pristine AA-BLG has a linear gapless energy spectrum which has attracted considerable theoretical interest\cite{Rakhmanov_2012,Mohammadi_2015,Chen_2013,Chiu_2013,Rozhkov_2011,Rozhkov_2016}. Different investigations on AA-BLG have been performed such as spin Hall effect\cite{Dyrda__2014,Hsu_2010},  doping effects\cite{Sboychakov_2013}, dynamical conductivity\cite{Tabert2012}, tunneling through electrostatic and magnetic barriers\cite{Wang_2013,Sanderson_2013}, magnon transport\cite{Owerre01_2016},  and the influence of spin orbit coupling on the band structure\cite{Yao_2007}. A recent work also studied the tunneling through an array of electrostatic barriers considering a mass term in the system without a bias\cite{Redouani2016}.

Because of the so-called Klein tunneling of Dirac fermions there is no complete confinement in graphene. One way to overcome this problem  by introducing a band gap in the energy spectrum which can be achieved, for example,  using slow Li$^+$ ions or perpendicular electric field in single-layer graphene and AB-BLG, respectively\cite{Oostinga_2007,Ohta_2006,Zhou_2007,Ryu_2016}.
Additionally, it has been observed that substrates can also play a major role in the electronic confinement of single layer graphene due to the substrate-induced band gap of the order $\backsim (20-500)$ meV\cite{Wang_2015,San_Jose_2014,Kindermann_2012,Song_2013,Jung_2015,Nevius_2015,Zarenia_2012}. The width of the band gap depends on the mass term induced by the substrates whose magnitude can be of the order $\backsim(50-100)$ meV depending on the type of the substrate \cite{Uchoa_2015}. Recently, a study showed that Hall phase can be realized in gapped AB-BLG when mass terms are considered in both layers\cite{Zhai2016}. Such mass terms are induced by dielectric materials such as hexagonal boron nitride (h-BN) or SiC.

 \begin{figure}[t!]
\vspace{0.cm}
\centering\graphicspath{{./Figures/}}
\includegraphics[width=2.8  in]{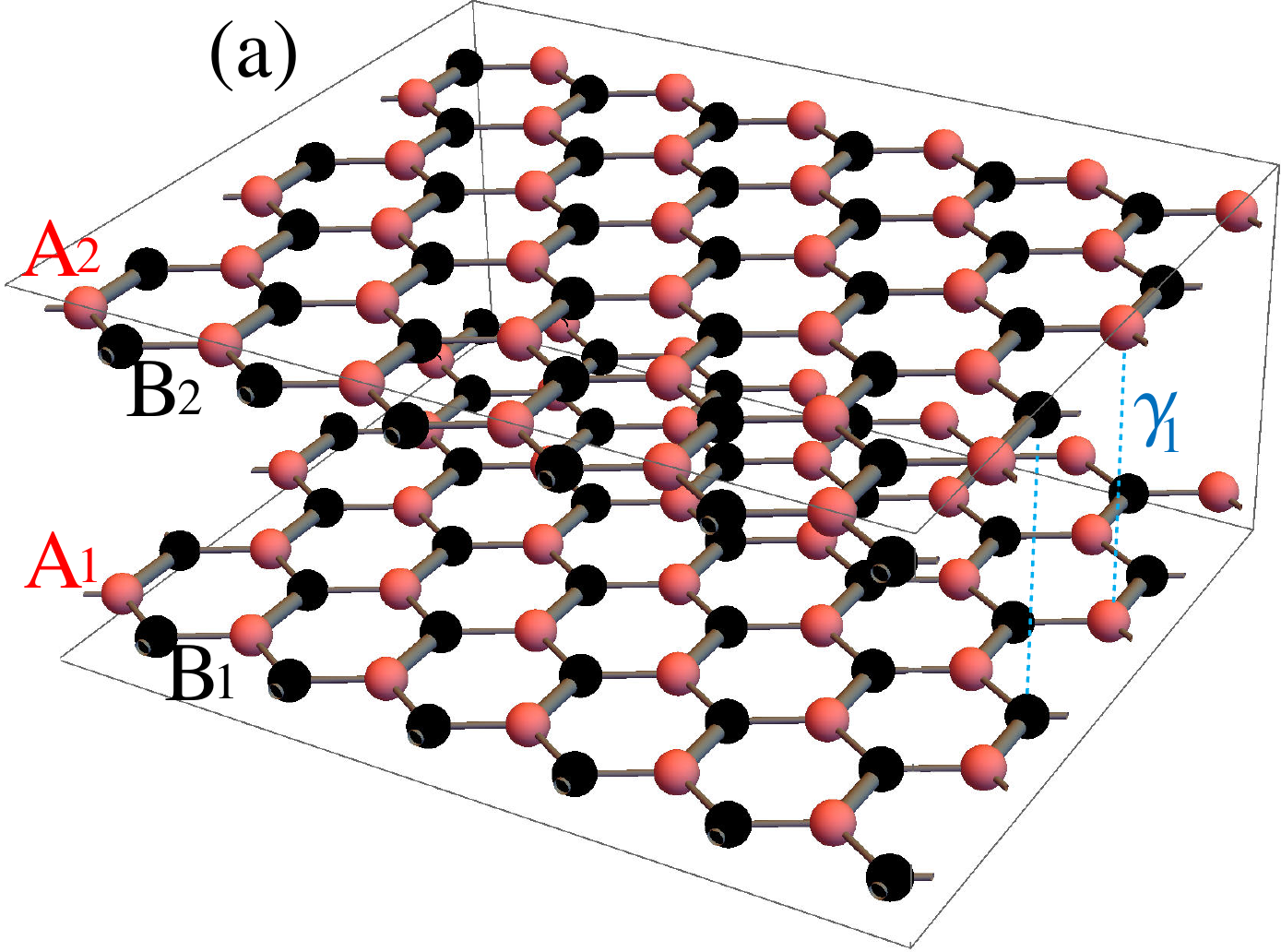}\\
\includegraphics[width=1.55  in]{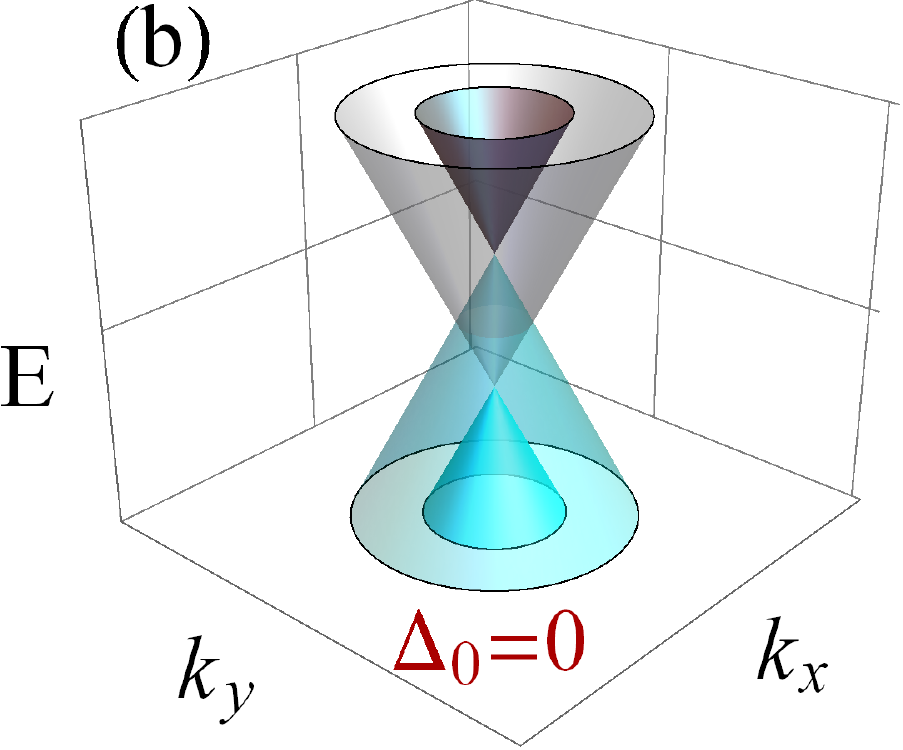}\ \ \ \ \ \
\includegraphics[width=1.55  in]{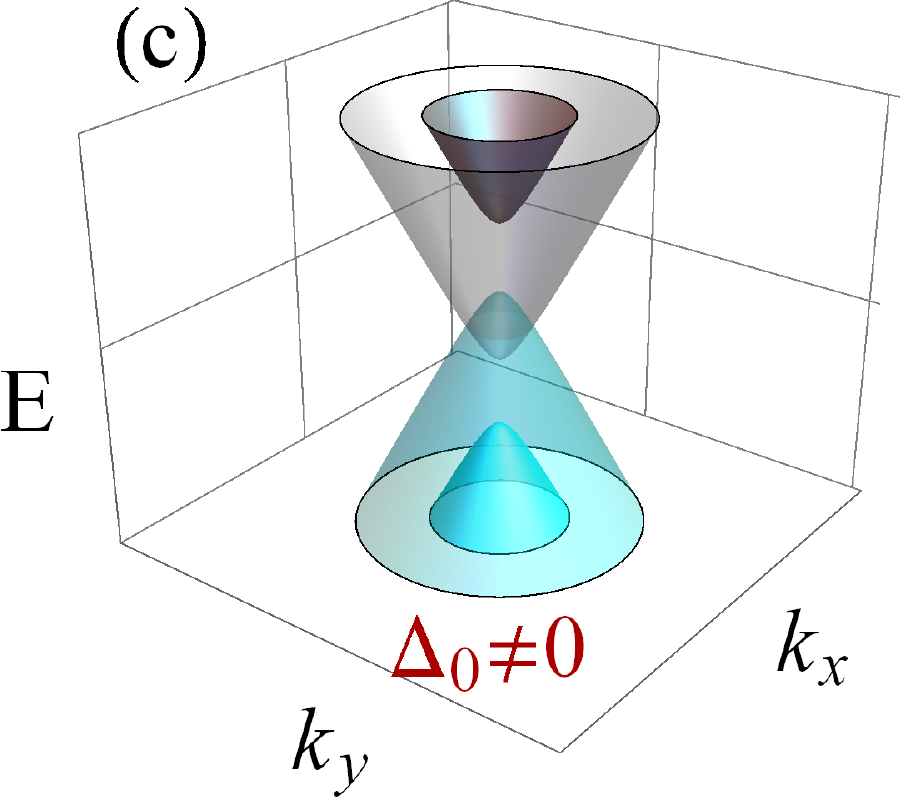}\ \ \ \ \ \ \
\vspace{0.cm}
\caption{(a) Crystalline structure of the AA-stacked graphene bilayer system and the associated energy spectrum in case of:  (b) zero and (c)  finite mass-term.   }\label{crys_structure}
\end{figure}

In the present work we investigate the substrate effect on intra- and inter-cone transport of biased AA-BLG. Considering the mass term to be the same in both layers of AA-BLG leads to  opening  a gap around the lower and upper cones whereas the whole spectrum remains gapless.  The presence of a  mass term can greatly affect the intra-cone transport. However, the total conductance of the system  remains almost unchanged.  Biasing the two layers of AA-BLG allows inter-cone transition due to coupling  the upper and lower cones. We also found that including a mass term in the system can significantly attenuate Klein tunneling in AA-BLG systems.

\section{Electronic Model}\label{Sec:Model}
\begin{figure}[t!]
\vspace{0.cm}
\centering\graphicspath{{./Figures/}}
\includegraphics[width=3.3  in]{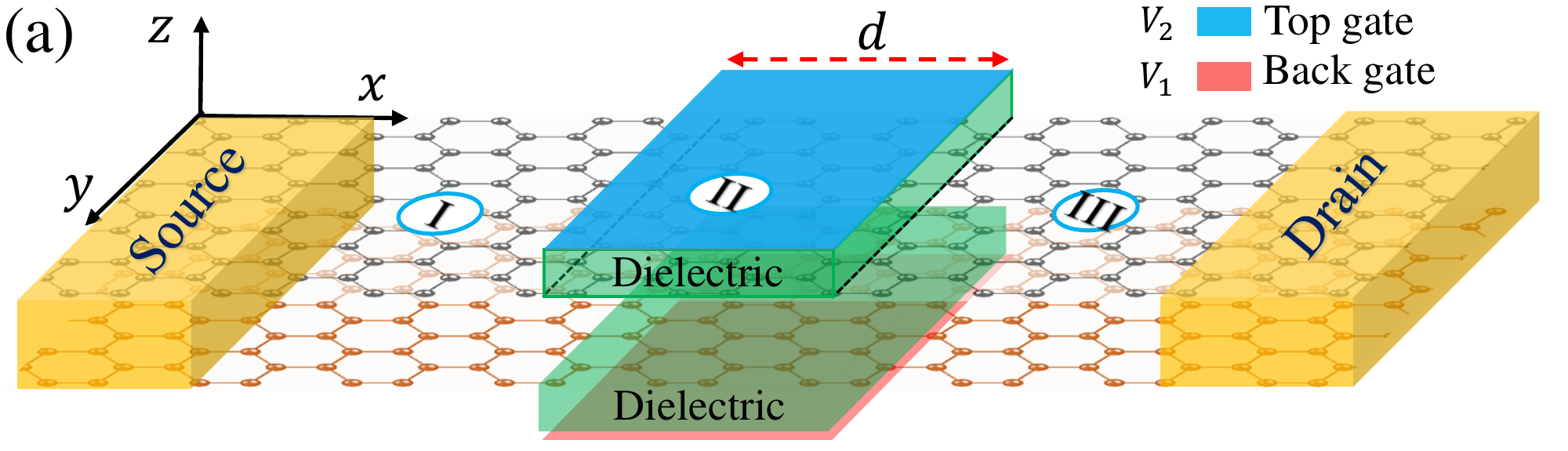}\\
\includegraphics[width=3  in]{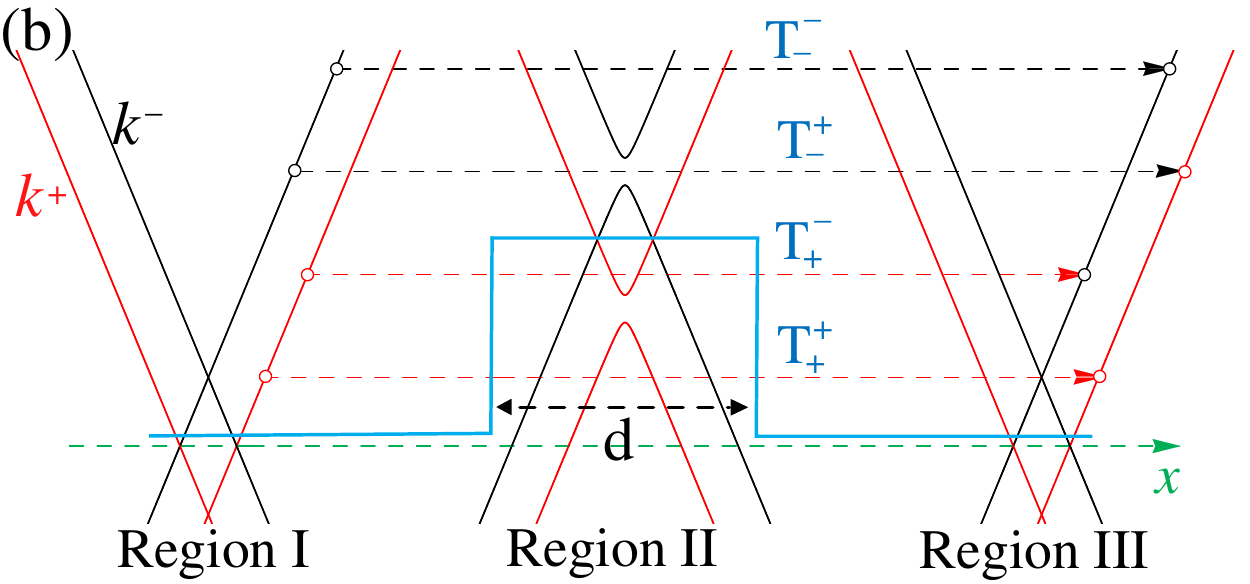}\ \ \ \ \ \ \ \  \ \ \

\vspace{0.cm}
\caption{  Schematic diagram  for the AA-BLG system subjected to top and bottom voltage gates which can control the electrostatic potential strength  $v_0$ and the inter-layer potential difference $\delta$. The  amplitude of the induced mass term is an inherent  character  associated with a certain  substrate and cannot be varied  externally. (b) Different possible intra- and inter-cone transitions through the junction. }\label{device}
\end{figure}

Single layer graphene has a hexagonal crystal structure and comprises
two atoms $A$ and $B$ in its unit cell whose interatomic distance is $a=0.142$ nm and its intra-layer coupling is $\gamma_0=3$ eV\cite{Zhang_2011}. In the AA-stacked graphene bilayer the two single layer graphene are placed exactly on top of each other such that  atoms $A_2$ and $B_2$ in the top layer are located directly above the $A_1$ and $B_1$ atoms in the bottom layer with a direct inter-layer coupling $\gamma_1 \approx 0.2\ eV$ \cite{Lobato_2011}, see Fig. \ref{crys_structure}(a). The energy spectrum with and without the mass term are shown in Figs. \ref{crys_structure}(b, c), respectively.  We see  that the mass term introduces a gap in the vicinity of the upper and lower cones. Notice that  the whole spectrum remains gapless unless the mass-term amplitude exceeds the inter-layer coupling.
The  continuum approximation of the Hamiltonian which describing electrons near the Dirac point $K$ in AA-BLG taking into account the mass term reads\cite{Tabert2012}
\begin{widetext}
\begin{equation}\label{AA_Hamiltonian}
H =\left(
\begin{array}{cccc}
    v_{0}+\delta+\Delta_0  & v_{\rm F}\hat{\pi}_{+} &  \gamma_{1} & 0 \\
  v_{\rm F}\hat{\pi}_{-} &  v_{0}+\delta-\Delta_0  &  0 & \gamma_{1}\\
  \gamma_{1} &   0& v_{0}-  \delta+\Delta_0  & v_{\rm F}\hat{\pi}_{+} \\
  0 & \gamma_{1}& v_{\rm F}\hat{\pi}_{-} & v_{0}-  \delta-\Delta_0   \\
\end{array}%
\right),
\end{equation}
\end{widetext}

where $\hat{\pi}_{\pm} =\left( p_x\pm\tau p_y  \right)$ is the momentum operator and $p_{x,y}=-i\hbar \partial_{x,y}$. The substrate effect is represented by the mass-term  whose amplitude is $\Delta_0=(\Delta_2+\Delta_1)/2$  where $\Delta_{1,2} $ denotes the mass term in the first and second layer, respectively. Note that we here assume zero inter-layer mass-term difference which requires $\Delta_1=\Delta_2$.  $v_0=(v_{2}+v_{1})/2$ and $\delta=(v_{2}-v_{1})/2$ describe the electrostatic potential strength and the bias, respectively,  subjected to region $II$ whose width is $d$ and can be modulated by external voltage gates as shown in Fig. \ref{device}(a).  $v_i$ is the electrostatic potential on the $i$-th layer and $v_F \approx 10^6$m/s is the Fermi speed of  charge carriers in the graphene sheet. It is well known that  substrates induce a mass term of opposite sign on sublattices $A_i$ and $B_i$, respectively.  A simplification can be made to the Hamiltonian in Eq. \eqref{AA_Hamiltonian},  by applying unitary transformation that forms symmetric and anti-symmetric combination of the top and bottom layer. This results in a Hamiltonian in the basis $\mathbf{\Psi}=2^{-1/2}(\Psi_{A 2}+\Psi_{A 1},\Psi_{B 2}+\Psi_{B 1},\Psi_{A 2}-\Psi_{A 1},\Psi_{B 2}-\Psi_{B 1})^{T}$.

Introducing the length scale $l=\hbar v_{F}/\gamma_{1}$, which represents the inter-layer coupling length $l \approx 3.3$ nm, allows
us to define the following dimensionless quantities:
\begin{widetext}
\begin{align*}
E\rightarrow\frac{E}{\gamma_1},\ v_0\rightarrow\frac{v_0}{\gamma_1},\ \delta\rightarrow\frac{\delta}{\gamma_1},\ \Delta_0\rightarrow\frac{\Delta_0}{\gamma_1},\  k_y\rightarrow lk_y,\ \text{and}\ \vec r\rightarrow\frac{\vec r}{l}.
\end{align*}
\end{widetext}
As a result of the translational invariance along the $y$ direction, the momentum along this direction is a conserved quantity and, hence, the wavefunction in the new basis can be written as
\begin{equation}\label{eq03}
\mathbf{\Psi}(x,y)=e^{iyk_y}\left[\phi_{1},\phi_{2},\phi_{3},\phi_{4}\right]^{\dag},
\end{equation} where $\dag$ stands for transpose. From Schrodinger equation $H \Psi = E \Psi$, one obtains four coupled differential equations that can be reduced to a single second order differential equation for $\phi_{2}$ as follows

\begin{figure}[t!]
\vspace{0.cm}
\centering\graphicspath{{./Figures/}}
\includegraphics[width=3.3  in]{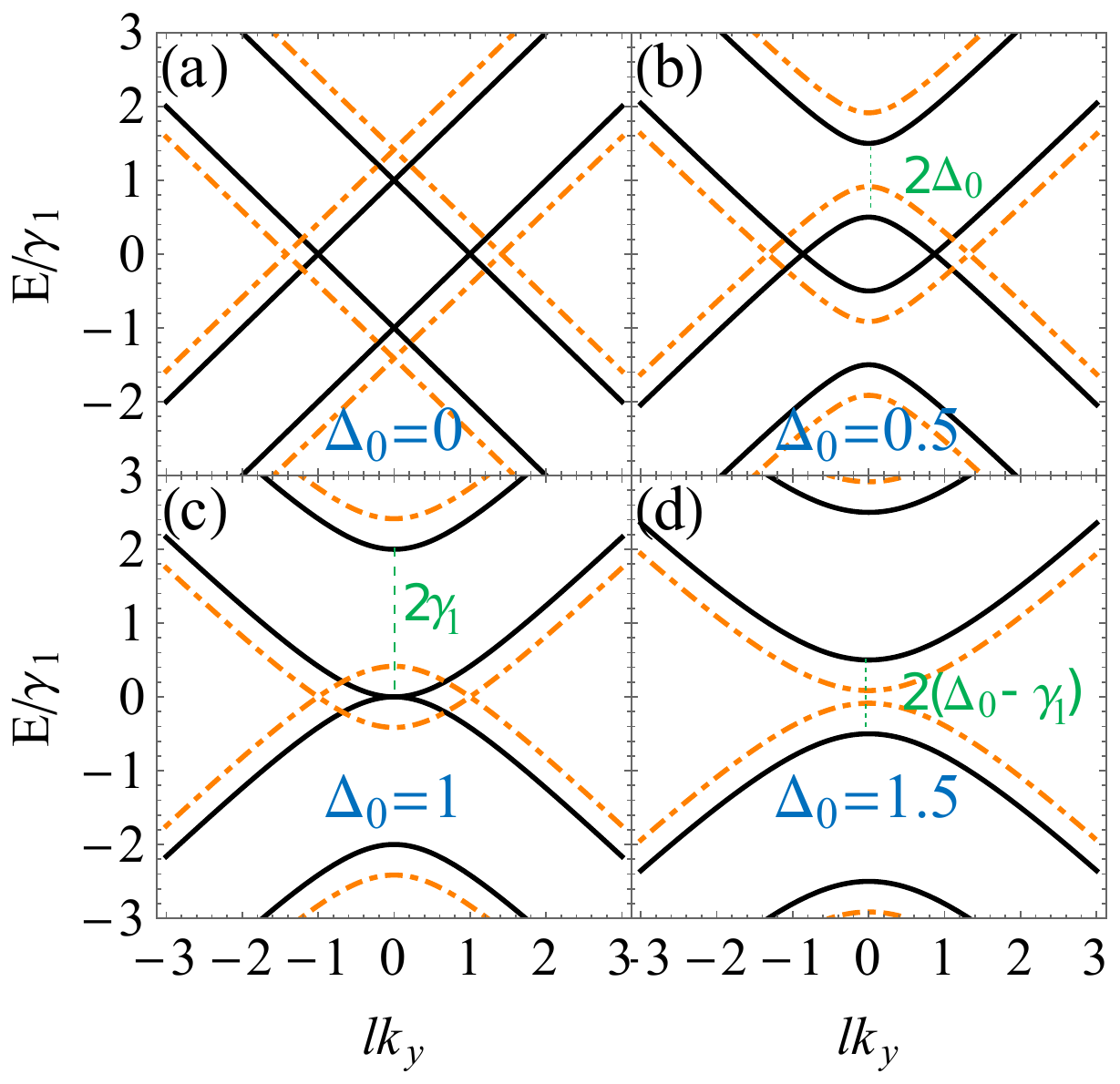}
\vspace{0.cm}
\caption{Energy spectrum of AA-stacked bilayer graphene  for $v_0=0$ and with (a)  zero mass, (b, c, d)   finite mass terms  that is smaller, same and larger  than the inter layer coupling $\gamma_1$, respectively.  Dashed orange and black curves correspond to the same system with ($\delta=\gamma_1$)   and without a bias, respectively. }\label{energy_spect}
 \end{figure}

\begin{equation}\label{decoupled_diff_Eq}
\left[\frac{d^{2}}{dx^{2}}+(k^{\pm}_x)^{2}\right]\phi_{2}=0,
\end{equation}
where
\begin{equation}\label{wave_vector}
k_x^{\pm}=\left[-k^{2}_{y}+\epsilon^2+\beta^{2}
\pm \sqrt{4\epsilon^2(1+\delta^2)}\right]^{1/2},
\end{equation}
with $\epsilon=E-v_0$
and $\beta^{2}=1+\delta^2-\Delta_0^2$ . The wave vectors $k_x^{\pm}$ and $k_y$ can be expressed in terms of the incident angle as
\begin{equation}\label{wav_vect}
k_x^{\pm}=(1\pm E)\text{cos}\phi,\ k_y= (1\pm E)\sin\phi.
\end{equation}
From Eq. \eqref{wave_vector}, it follows that the energy spectrum for the system is given by
\begin{equation}\label{eq10}
\epsilon_\xi^{\pm}=\xi\left[
k^{2}_{y}+\beta^{2}+2\Delta_0^2\pm2 \sqrt{(1+\delta^2)(k^{2}_{y}+\Delta_0^2)}\right]^{1/2},
\end{equation}
where $\xi=\pm1$. In Fig. \ref{energy_spect}. We show the energy spectrum of the AA-BLG for different values of the system parameters. Pristine
AA-BLG has a linear energy  spectrum  with two up-down Dirac cones shifted by $\Delta E$, which is $2\gamma_1$ in this case as shown in Fig. \ref{energy_spect}(a) by the solid black curves. When AA-BLG is subjected to a perpendicular electric field (biased AA-BLG) the two Dirac cones are slightly shifted and located at $v_0\pm\sqrt{\gamma_1^2+\delta^2}$ , see dotted-dashed orange curves in Fig. \ref{energy_spect}(a). Fig. \ref{energy_spect}(b) is the same as Fig. \ref{energy_spect}(a) but in the presence of mass-term amplitude $\Delta_0=0.5\gamma_1$ and the spectrum exhibits a shift $\Delta E_m=2\Delta_0$ in the bands in the vicinity of the upper and lower Dirac cones. Note that the whole spectrum remains gapless.  When the mass-term amplitude is the same as the interlayer coupling, i.e. $\Delta_0=\gamma_1$, the spectrum becomes parabolic and resembles that of pristine AB-stacked BLG. The two lower bands touch each other at $E=0$ while the other two are spaced from the zero energy by $2\gamma_1$, see Fig. \ref{energy_spect}(c).  Considering the mass-term amplitude that exceeds the inter-layer coupling results in opening a direct gap in the energy spectrum of  magnitude $2(\Delta_0-\gamma_1)$ as shown in Fig. \ref{energy_spect}(d).

To calculate the transmission probabilities, the desired solution in each region must be obtained. Then, implementing the transfer matrix together with appropriate boundary conditions gives the transmission and reflection probabilities\cite{Barbier01_2010,Van_Duppen01_2013,Abdullah_2017}.  The different possible intra- or inter-cone transitions processes through the junctions are described in Fig. \ref{device}(b). The zero temperature conductance can be calculated using the B\"uttiker's formula\cite{Snyman_2007,Blanter_2000}  \begin{equation}
 {G_i^{j}}(E)=G_{0}\frac{L_y}{2
\pi}\int_{-\infty}^{+\infty}dk_{y} T_{i}^j(E,k_y),
\end{equation}
with $(i,j)=\pm$, $T^j_i$ represents the transmission probability of a particle  incident from the mode $k^i$(subscript $i$ in $T^j_i$) and transmitted to the mode $k^j$(superscript $j$ in $T^j_i$), while   $L_y$  defines the length of the sample in the $y$-direction, and
$G_0=4\ e^2/h$. The factor $4$ comes from the valley and
spin degeneracy in graphene.
The total conductance of the system is the sum through all available channels.

\section{Results and discussion}\label{Sec:Results}
\begin{figure}[t!]
\vspace{0.cm}
\centering\graphicspath{{./Figures/}}
\includegraphics[width=3.3  in]{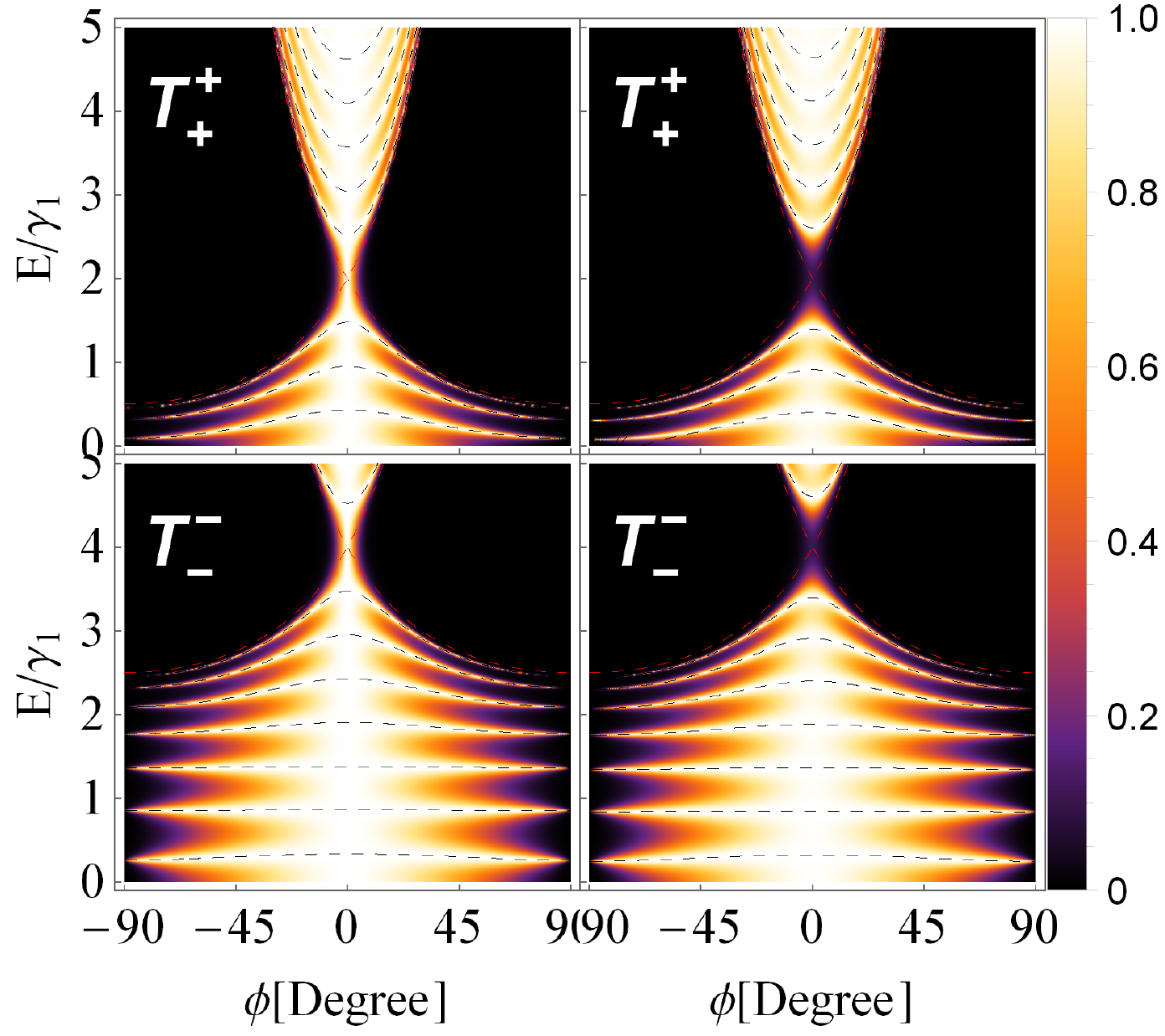}\\
\vspace{0.cm}
\caption{Density plot of the transmission probabilities for the lower $ T_+^+ $ and upper $ T_-^- $ Dirac cones with $v_0=3\gamma_1$, $d=6l$ and $ \delta=0$.  Left and right panels for  $ \Delta_0=0$ and $ \Delta_0=0.3\gamma_1$. The superimposed  dashed white curves correspond to resonances resulting from the finite-size effect found using Eq. \ref{resonances}. The red dashed bands correspond to   AA-BLG with $v_0=3\gamma_1$ in the intermediate region II.  The scattered channels $T_+^-$ and $T_-^+$ are zero since the inter-layer electrostatic difference is zero i.e. $\delta=0.$ }\label{T_0_S}
\end{figure}

\begin{figure}[t!]
\vspace{0.cm}
\centering\graphicspath{{./Figures/}}
\includegraphics[width=3.  in]{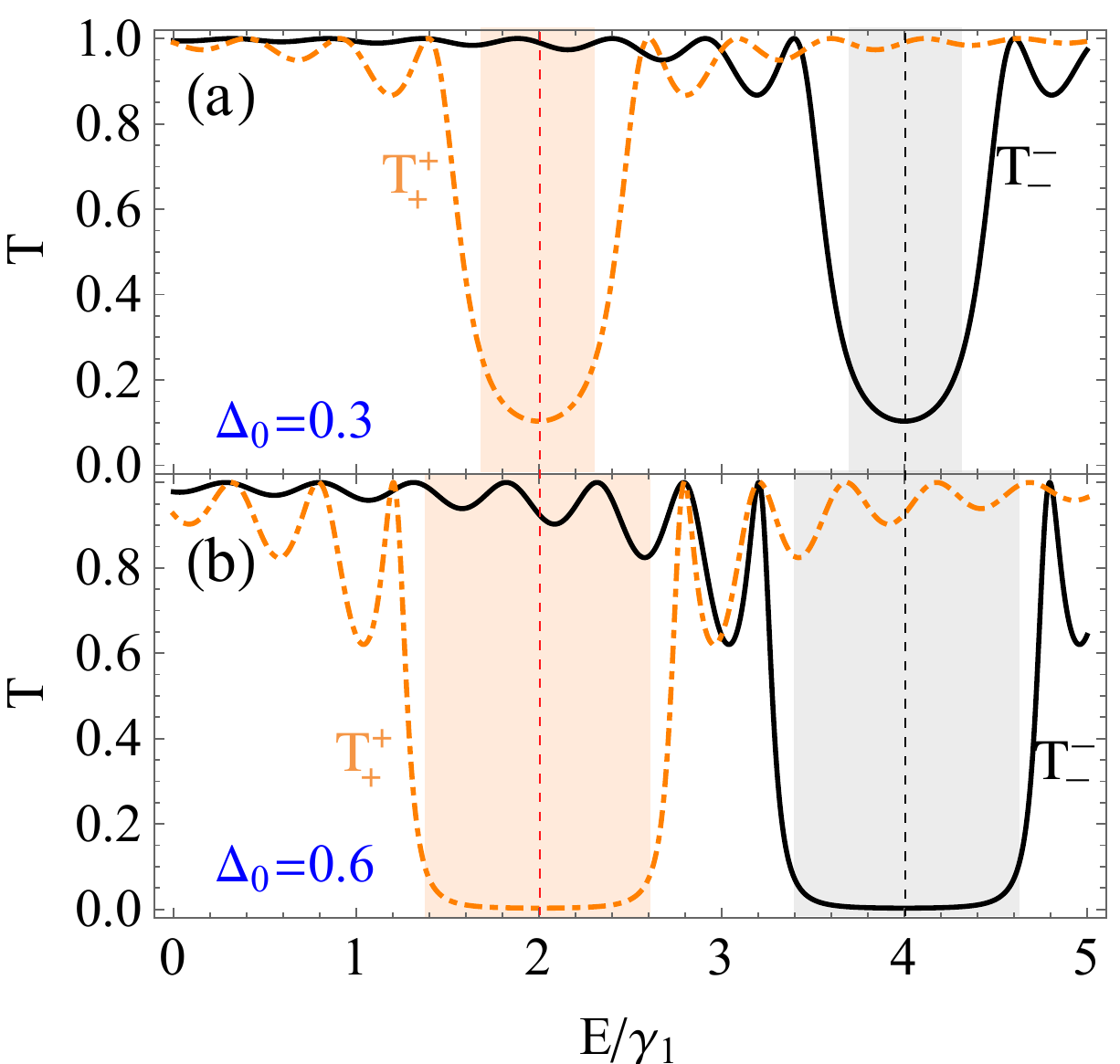}\\
\vspace{0.cm}
\caption{Transmission probabilities of the lower and upper cones for normal incidence, i.e.  $\phi=0$, with different mass-term amplitude. In both panels the bias $\delta$ considered to be zero while the potential strength $v_0$ is $3\gamma_1$. The vertical dashed red and black lines correspond to the position of the lower and upper cones in case of zero mass-term amplitude.  While the red and black dashed region represent the opened gap in the vicinity of the lower and upper cones. respectively.  }\label{norm_inc}
\end{figure}

\begin{figure}[t!]
\vspace{0.cm}
\centering\graphicspath{{./Figures/}}
\includegraphics[width=3.3  in]{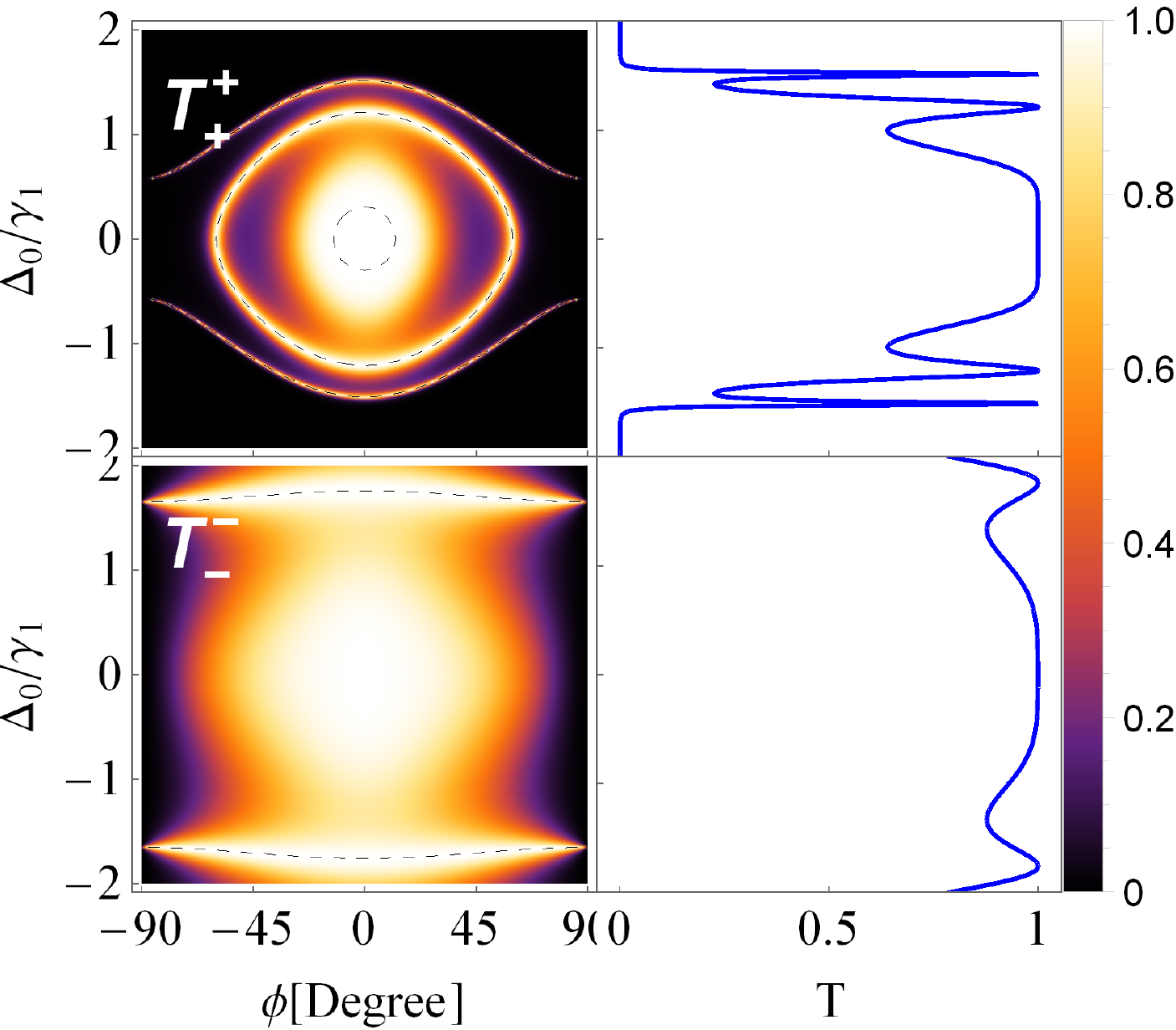}\\
\vspace{0.cm}
\caption{(Left panel) Density plot of the transmission probabilities as a function of the mass-term amplitude and the incident angle for the lower $T_+^+$ and upper $T_-^-$ Dirac cones with $v_0=3\gamma_1$, $d=6l,\ \epsilon=0.4\gamma_1$ and $\delta=0$.  (Right panel) Transmission probabilities along the normal incidence direction associated with each cone, the dashed black curves represent the    Febry-P\'erot resonances. }\label{mass_dep}
\end{figure}

\begin{figure}[t!]
\vspace{0.cm}
\centering\graphicspath{{./Figures/}}
\includegraphics[width=3.3  in]{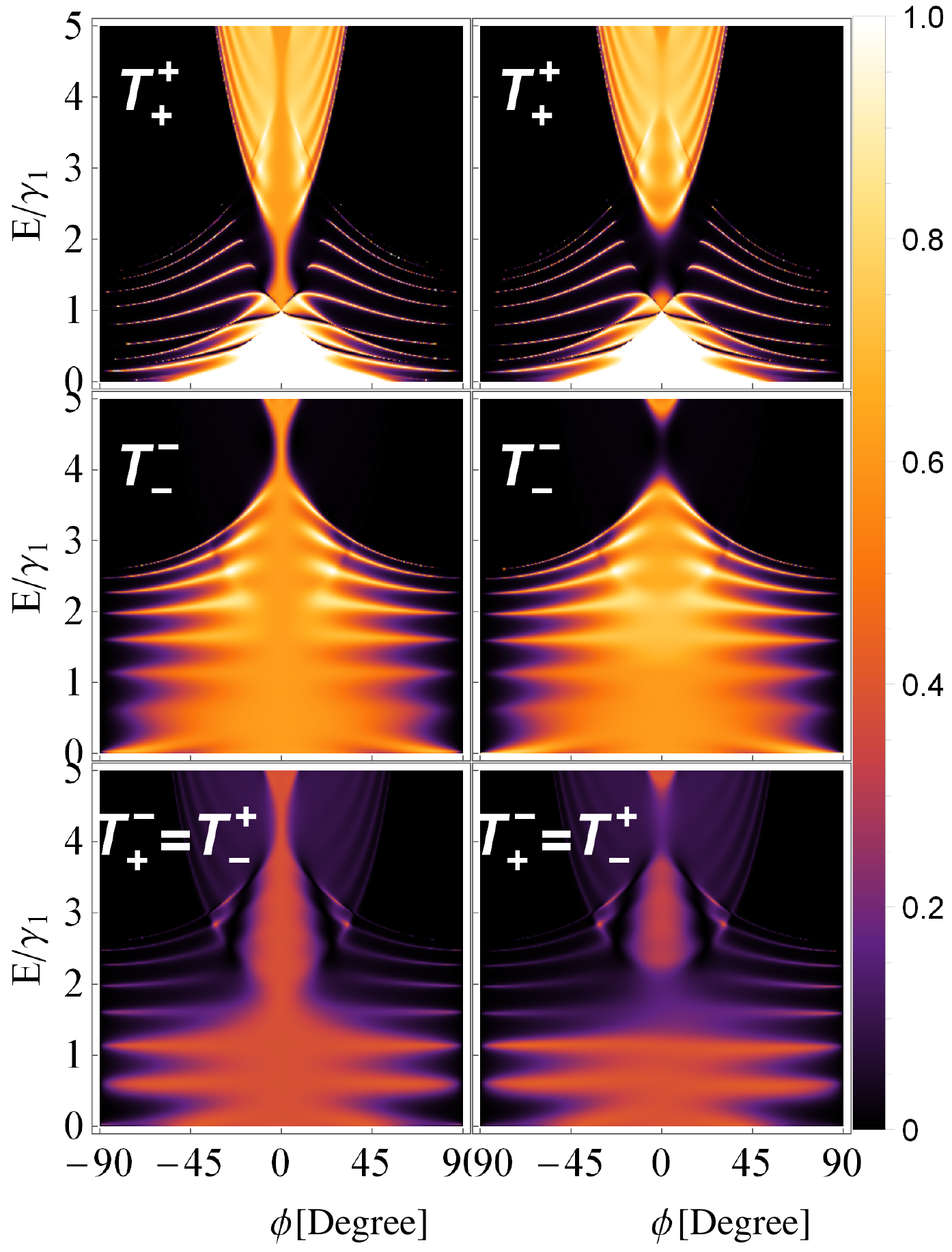}
\vspace{0.cm}
\caption{Same as in Fig. \ref{T_0_S}, but now with finite bias $\delta=0.8\gamma_1$. Note that, in contrast to the previous case, the bias here induces the  scattering between the two Dirac cones such that  $T_+^-=T_-^+\neq0$ , hence inter-cone transport is switch on.   }\label{T_v_S}
\end{figure}
\begin{figure}[t!]
\vspace{0.cm}
\centering\graphicspath{{./Figures/}}
\includegraphics[width=3.3  in]{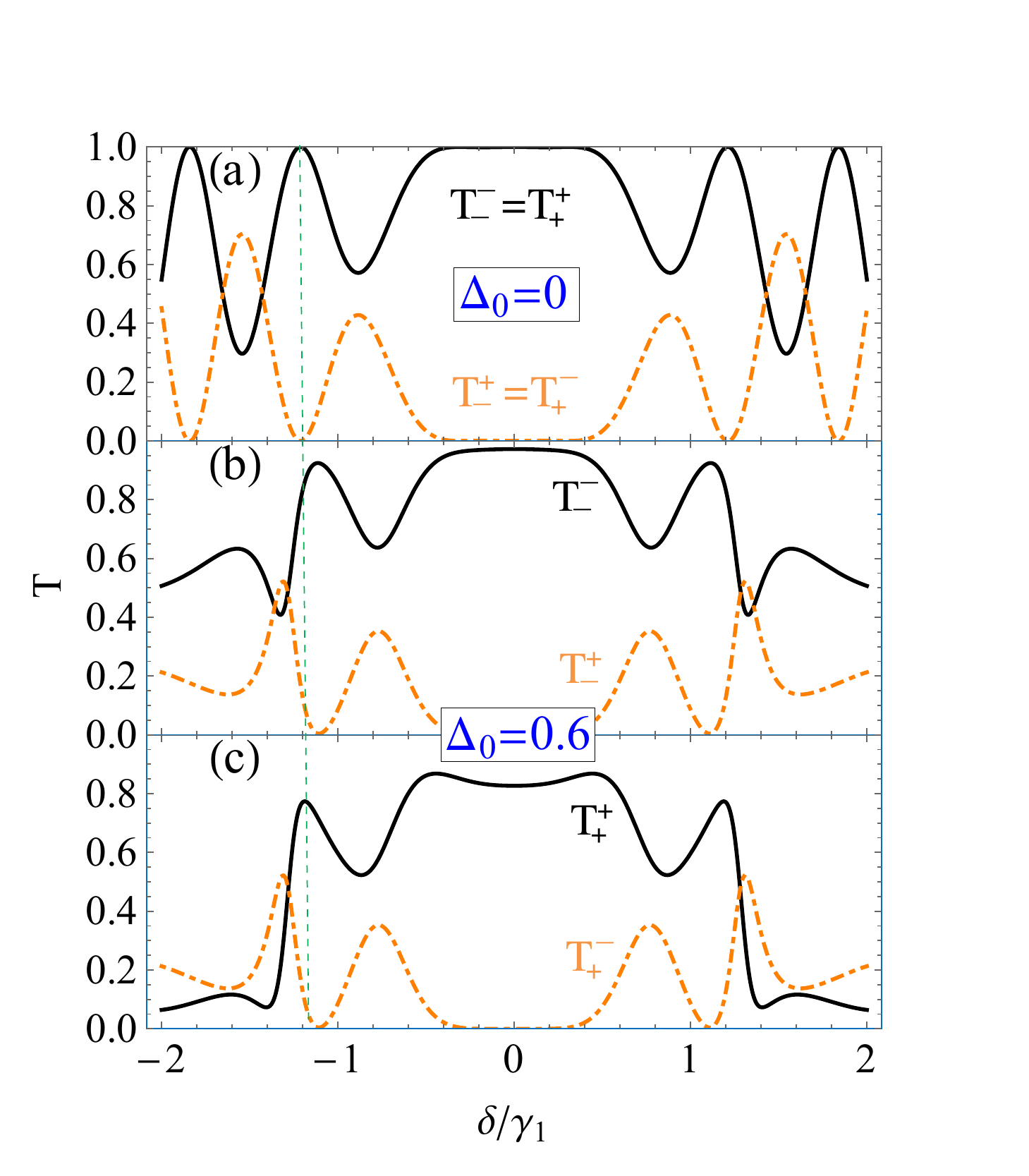}
\vspace{0.cm}
\caption{ Transmission probabilities of the intra- (solid black) and inter- (dashed orange) cone channels  as a function of the bias for normal incidence $\phi=0$ with $\epsilon=0.6\gamma_1$. (a) Zero and  (b, c) finite  mass-term amplitude. Note that, for normal incidence and in case of zero mass term in panel (a) the non-scattered channels  are the same $T_-^-=T_+^+$ as well as   scattered ones such that $T_-^+=T_+^-$. The vertical dashed green line is located at a point  with    perfect  tunneling in  panel (a) associated with intra-cone channel.}\label{v_dep}
\end{figure}
In this section, we first analyze the Fermi   energy-dependent behavior of intra- and inter-cone transport. In the absence of the bias only the intra-cone transport is allowed which  conducts through the channels $T_+^+$  and $T_-^-$ corresponding to the lower and upper cones, respectively.
In Fig. \eqref{T_0_S} we show the transmission probabilities for $\delta=0$ and $v_0=3\gamma_1$ with $\Delta_0=0(0.3)\gamma_1$ in the left(right) panel.  In the absence of the mass-term (i.e. $\Delta_0=0$ ) and for normal incidence (i.e. $\phi=0$ )  the lower  and upper  cone  transmission probabilities show perfect  tunneling  even when the
barrier strength is larger than the Fermi energy,  see Fig. \ref{T_0_S}(left panel).  In other words, $T_+^+(\epsilon,0) = T_-^-(\epsilon,0)=1,$ 
such behaviour  is a typical signature of Klein tunneling in graphene. However, introducing the mass-term (i.e. $\Delta_0\neq0$) results in a significant suppression in the  transmission for  specific ranges of energy, specifically, in vicinity of lower and upper cones, see Fig. \ref{T_0_S}(right panel). For the lower (upper) cone channel $T_{\mp}^{\mp}$, the suppression  occurs for $v_0\pm\gamma_1+\Delta_0\geqslant E\geqslant v_0\pm\gamma_1-\Delta_0$. Notice that the total charge carrier transmission of the system is non-zero in these energy ranges. To examine thoroughly  the effect of  substrate on  Klein tunneling, we show in Fig. \ref{norm_inc} the transmission probabilities of the lower and upper cones for normal incidence.
In panel (a) we assume that the substrate induces a mass-term  of magnitude $0.3\gamma_1$ while it is doubled in  panel (b).  We see that the transmission probabilities display different behaviours in this case. Of particular importance we notice that the Klein tunneling is attenuated and rather resonances appear
in the transmission probabilities as a results of the finite size effect. These so called Febry-P\'erot resonances\cite{Snyman_2007} coincide with the energies given by
\begin{equation}\label{resonances}
\epsilon_{\alpha,n}^{\pm}=-\alpha+
\frac{v_0\pm \sqrt{v_0^2\sin^2\phi+\left[\left( \frac{n\pi}{d} \right)^2+\Delta_0^2\right]\cos^2\phi}}{\cos^2\phi}
\end{equation}
where $\alpha=(+,-)$ is the cone index associated with (lower, upper) cones  and $\pm$ stands for electron- and hole-like states corresponding to each cone. We superimpose these resonant energies as dashed black curves  in Fig. \ref{T_0_S}. Note that equation \eqref{resonances} is valid for any angle in contrast to the AB-BLG where its validity only holds for normal incidence. We know that, for normal incidence, the presence of Klein and anti Klein tunnelling  in single and AB-BLG, respectively, is a direct consequence of the pseudo spin conservation in the system. This means that in AA-BLG the backscattering is not forbidden and a non-zero reflected current can appear at certain energies when considering  a  mass term. Moreover, the intra-cone transport  along the normal incident direction shows a rapid decrease in the vicinity of both cones (inside the barrier, i.e. $E=v_0\pm\gamma_1$ ), see Fig. \ref{norm_inc}(a). This decrease can reach zero when a large mass-term  amplitude is considered as shown in Fig. \ref{norm_inc}(b) . In Fig. \ref{mass_dep} we show the intra-cone transmission as a function of the mass-term amplitude and the incident angle for a certain energy. For a specific energy, the role of the mass term  in the lower and upper cone transmission are distinct as reflected in Fig. \ref{mass_dep}(left panel). The lower cone transmission channel $T_+^+$ is completely suppressed for either large mass-term amplitude or large incident angle. While the charge carriers can still conduct for wide ranges of energy and mass-term amplitudes through the upper cone channel $T_-^-$.  In Fig. \ref{mass_dep}(right panel) we show  the transmission probabilities, associated with each cone,  along  the normal incidence direction which clarify the suppression in lower cone transmission for large  $\Delta_0$.

The main role of the bias $\delta$ in the cone-transport  can be deduced from the Hamiltonian in Eq. \eqref{AA_Hamiltonian}, after performing a unitary transformation, where it couples the lower and upper Dirac cones  leading to the inter-cone transport\cite{Abdullah2017}. To investigate the inter-cone transmission induced by the bias ($\delta\neq0$) we show the density plot for the different channels for $\Delta_0=0\ (0.3)\gamma_1$  in the left (right) panel of Fig. \ref{T_v_S}. The transmission profile in Fig. \ref{T_v_S}(bottom panels) shows clearly that, due to the bias, scattering between the upper and lower cones $T_+^-(T_-^+)$ is significant. For comparison with the results of a biased AB-BLG\cite{Van_Duppen01_2013}, it can be seen  here that  the scattering channels $T_+^-(T_-^+)$ preserve the angular symmetry in contrast to the case of AB-BLG where such symmetry is broken. This difference is a manifestation of the symmetric and asymmetric inter-layer coupling in the AA-BLG and AB-BLG, respectively.
The fringes that appear in $T_+^+$ are also a consequence of the bias and their number depends on the width of the intermediate region. Such fringes arise due to the well known Fabry-P\'erot interference between the two coupled modes. These fringes are not affected when a mass term is considered as can be inferred from the channel $T_+^+$ in Fig. \ref{T_v_S}(right panel). In contrast to the intra-cone transport, it is observed that regardless  amplitude strength of  the mass term, the inter-cone transport channels $T_+^-(T_-^+)$ show a significant transmission in the vicinity of lower and upper cones (inside the barrier) as depicted in Fig. \ref{T_v_S}(right panel). This can be understood from  the bands  in Fig. \ref{device}(b). For intra-cone  channels, for instance $T_+^+$, around the lower cone there are $k^+$(red band in Fig. \ref{device}(b)) propagating states outside the barrier but they are absent inside the barrier within the same range of energy. Hence, a gap in the intra-cone channels arises in the vicinity of lower and upper cones. On the other hand, the non-zero transmission associated with the inter-cone channels in the vicinity of the Dirac cones  is a manifestation of the availability of states with opposite parity inside the barrier. For example, in the channel $T_+^-$ the incident charge carriers coincide with the $k^+$ mode and jump to states whose mode is of opposite parity, i.e. $k^-$.  The presence of the $k^-$ inside the barrier in the vicinity of the lower cone, inside the barrier, leads to a significant tunneling in the channel $T_+^-$.  The same analogy  also applies to the channel $T_-^+$.
\begin{figure}[t!]
\vspace{0.cm}
\centering\graphicspath{{./Figures/}}
\includegraphics[width=3.  in]{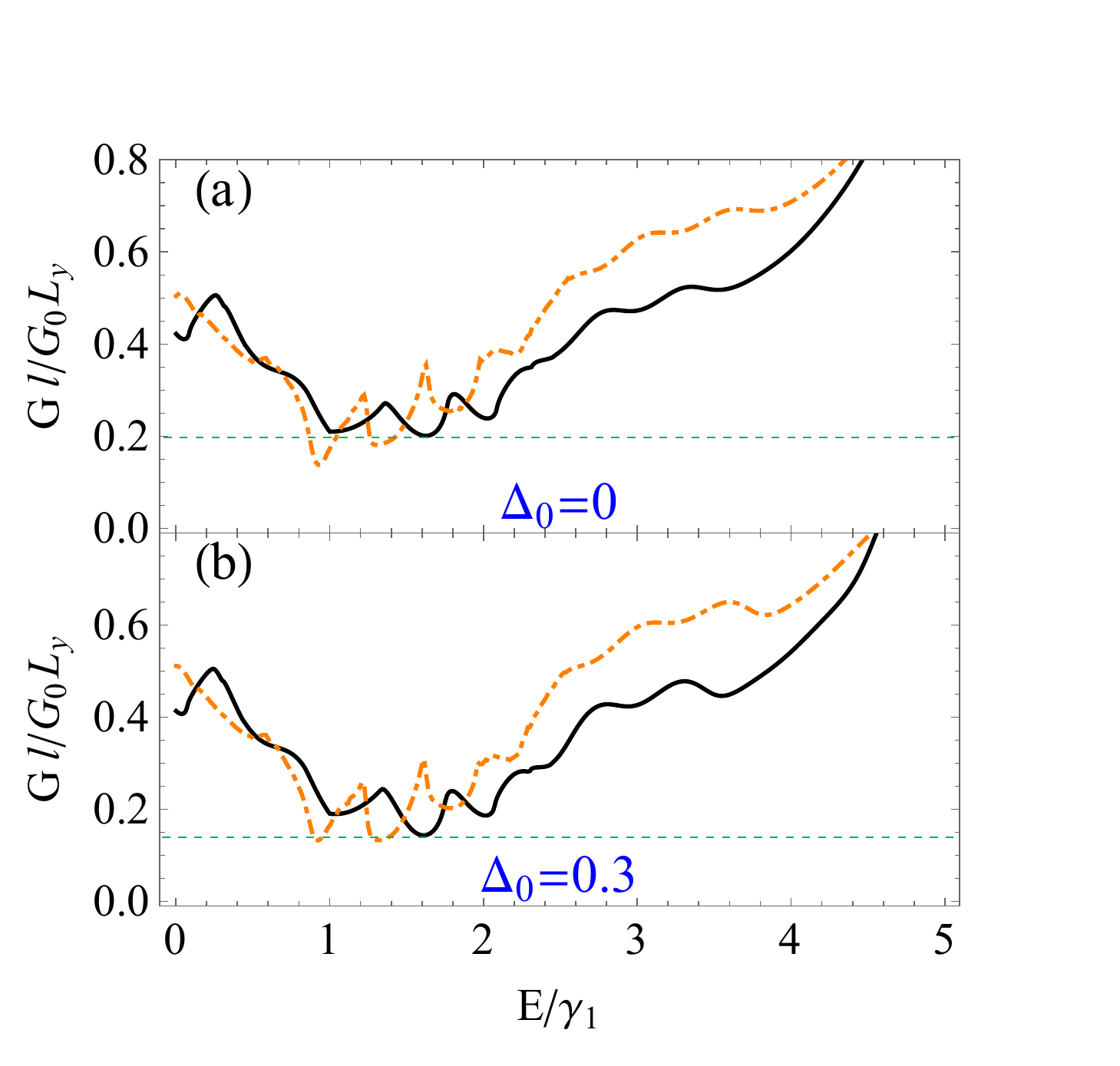}
\vspace{0.cm}
\caption{ Total conductance as a function of  Fermi energy with zero (a) and finite (b)  mass-term amplitude. The solid black and dashed brown curves correspond to the system with ($\delta=0.5\gamma_1$) and without a bias, respectively. }\label{cond}
\end{figure}

In Fig. \ref{v_dep}, we show the intra- (solid black) and inter- (dashed brown) cone transport along the normal incidence direction as a function of the bias for a certain energy. The intra- and inter-cone transport show a certain complementarity such that  a maximum in first one   coincide with minimum in the latter one as can be seen in Fig. \ref{v_dep}(a). Such behaviour is a manifestation of  Klein tunneling being preserved which enforces the total transmission of each cone to reach unity at normal incidence. This means that for the lower cone  we need  $T_+^++T_+^-=1$ and for the upper cone $T_-^-+T_-^+=1$. This condition leads to the equivalence in the intra- and inter-cone  channels such that $T_-^-=T_+^+$  and $T_+^-=T_-^+$. Considering the mass term results in an attenuated Klein tunneling and hence breaking the equivalence in the intra-cone channels $T_+^+$ and $T_-^-$ as can be inferred from Figs. \ref{v_dep}(b, c) (see solid black curves).  However, the equivalence in the inter-cone channels is still obtained as can be seen in Figs. \ref{v_dep}(b, c) (brown dashed curves) which is a result of the valley degeneracy\cite{Van_Duppen01_2013}.  Note that  the observed complementarity in  the intra- and inter-cone channels is also lost as shown in Figs. \ref{v_dep}(b, c).  

The total conductance of the system as a function of the Fermi energy is shown in Figs. \ref{cond} (a,b) for zero and finite mass-term amplitude, respectively. Without bias and by comparing solid black curves in panels (a) and (b),  we find that the  behaviour of the total conductance remains almost the same.
This means that even though the mass term has a significant impact on the intra-cone transport, the total conductance remains almost unchanged when considering the small mass-term amplitude. Note that these two channels, $T_+^+$ and $T_-^-$, cannot be distinguished  experimentally. In other words,  the measurements will give the total  transport of both channels and as a result  the effect of the mass term on individual   channel cannot be observed.    However, the presence of the mass term reduces the overall conductance as indicated by the horizontal dashed green lines in Figs. \ref{cond}(a,b).
In contrast, biasing the system ($\delta=0.5\gamma_1$)\cite{Zhou2006} leads to a considerable increase in the overall conductance  as clarified by the brown dashed curves in Fig.  \ref{cond}(a,b).
We note also that the  conductance gets very pronounced as a result of the extra  inter-cone channels activated by the bias. The extra peaks appearing in the profile of the
conductance with  bias  are attributed to the fringes in the lower cone channel (see $T_+^+$ in Fig. \ref{T_v_S}) as a result of the finite-size effect.

\section{Summary}\label{Sec:Summary}

In summary,  we have theoretically investigated the quantum transport across a biased AA-BLG $n$-$p$-$n$ junction. In the presence of an induced mass-term, the intra-cone transport is almost completely suppressed in the vicinity of the upper and lower cones. However, the system remains gapless and the behaviour of the total conductance remains almost the same when considering small mass-term amplitudes. The results also show that the inter-cone transport can only be induced and modulated by introducing a bias. As a result of the induced inter-cone transport, extra resonances appear in transmission profiles of the lower cone which appear in the evanescent modes regime of the upper cone. In general, the bias enhances the overall conductance in the presence or absence of the induced mass-term. Such increase in the conductance, especially for large energies, is attributed to  the additional inter-cone channels that can be accessed in the presence of a bias. In contrast to the normal incidence in  AA-BLG junction, where we observed an  attenuated Klein tunneling in the presence of the mass term.

It is our hope that the results in this article will provide a path for the electrical control of quantum transport in biased AA-BLG-based electronic device taking into account the substrate effect.

\section*{Acknowledgment}\label{Sec:Summary}
We acknowledge the support of King Fahd University of Petroleum and Minerals under research group project RG1502-1 \& RG1502-2. 
We also acknowledge the material support of the Saudi Center for Theoretical Physics (SCTP) during the progress of this work.
HMA thanks Mohammed M Al Ezzi for the critical readings of the manuscript.



\end{document}